\documentclass[aps,prd,preprint,onecolumn,superscriptaddress,nofootinbib]{revtex4}
\usepackage[T1]{fontenc}
\usepackage[utf8]{inputenc}
\usepackage{amsmath,amssymb,amsfonts,bm}
\usepackage{mathtools}
\usepackage{booktabs}
\usepackage{array}
\usepackage{hyperref}
\usepackage{xcolor}
\usepackage{orcidlink}
\usepackage{microtype}
\usepackage{enumitem}
\usepackage{xspace}

\hypersetup{colorlinks=true, linkcolor=blue!55!black, citecolor=blue!55!black, urlcolor=blue!55!black}

\allowdisplaybreaks

\newcommand{\mc}{\mathcal}

\newcommand{\KR}{Kalb-Ramond\xspace}

\newcommand{\Godel}{G\"odel\xspace}

\begin{document}

\title{\Godel-type universes in Lorentz-violating Kalb-Ramond gravity\\
with Ricci- and Riemann-tensor nonminimal couplings}

\author{Fernando M. Belchior\orcidlink{0009-0006-8675-7849}}
\email{fernandobelcks7@gmail.com}
\affiliation{Departamento de Física, Universidade Federal da Paraíba, Centro de Ciências Exatas e da Natureza, 58051-970, João Pessoa, Paraíba, Brazil}

\begin{abstract}
This paper studies homogeneous \Godel-type geometries in a gravitational model where local Lorentz symmetry is spontaneously broken by the vacuum expectation value of an antisymmetric Kalb-Ramond two-form. We consider the Einstein-Hilbert action supplemented by two independent nonminimal curvature couplings: a Ricci-tensor coupling of the form $B_{a}{}^{c}B_{bc}R^{ab}$ and a Riemann-tensor coupling of the form $B^{ab}B^{cd}R_{abcd}$. In the homogeneous vacuum sector, where the Kalb-Ramond field strength and the first derivatives of the symmetry-breaking potential vanish, the field equations reduce to a closed algebraic system in the tetrad frame of the \Godel-type metric. We derive the curvature tensors, the reduced metric equations, and the Kalb-Ramond consistency equations for several inequivalent orientations of the vacuum two-form. Particular attention is given to the pseudo-magnetic background $B_{12}=b$, because it preserves the symmetry of the rotation plane and yields a transparent deformation of the usual bumblebee result. For this sector the Kalb-Ramond equation fixes $\frac{m^2}{\omega^2}=\frac{2(\xi_1+3\xi_2)}{\xi_1+2\xi_2}$, where $\xi_1$ and $\xi_2$ are the Ricci and Riemann couplings. The Ricci coupling alone reproduces the noncausal \Godel value $m^2=2\omega^2$, while the Riemann coupling shifts the chronology bound and allows the critical causal value $m^2=4\omega^2$ for $\xi_2=-\xi_1$. We also show that the Riemann coupling can move the critical radius for closed timelike curves, but ordinary positive-energy matter severely restricts completely causal solutions. Finally, we discuss perfect fluid, scalar field, and electromagnetic sources. 
\end{abstract}

\maketitle

\section{Introduction}

The original solution found by \Godel is one of the most instructive exact space-times of general relativity \cite{Godel:1949,HawkingEllis,ReboucasTiomno:1983,ReboucasSantos:1983, BampiZordan:1978, Novello:1983}. It is homogeneous, rotating, regular, and sourced by physically simple matter, yet it contains closed timelike curves. The example remains important because it exposes a conceptual feature of Einstein gravity that is not visible in weak-field situations, where local causal cones can be perfectly well behaved while the global causal structure of the manifold permits chronology violation \cite{Accioly:2002, FonsecaNeto:2013}. This is why \Godel-type metrics have become a useful testing ground for modified theories of gravity. A theory that changes the effective relation between curvature, rotation, and matter can either preserve the usual acausal region, shift the critical radius for the formation of \Godel circles, or forbid such solutions for a given class of sources. Other works addressed the study of \Godel-types universes in modified gravity scenarios \cite{Furtado:2010, Porfirio:2016ssx, Furtado:2019, Nascimento:2020jwx, Nascimento:2021bzb, Nascimento:2023}

A particularly natural setting in which to revisit this issue is gravitational Lorentz violation \cite{KosteleckySamuel:1989, ColladayKostelecky:1997, ColladayKostelecky:1998,Kostelecky:2004}. In effective field theory, Lorentz violation can arise when a tensor field acquires a nonzero vacuum expectation value. The underlying action may still be observer covariant, but the vacuum selects preferred directions in the local Lorentz frame. This distinction is important in gravity. Explicit breaking is generically in tension with the geometrical identities of Riemann-Cartan geometry, while spontaneous breaking can be made compatible with the Bianchi identities because the coefficient for Lorentz violation is itself a dynamical field. Bumblebee models, in which a vector field develops a nonzero vacuum value, provide the simplest illustration \cite{BluhmKostelecky:2005,KosteleckyPotting:2009, BaileyKostelecky:2006, KosteleckyTasson:2011, Bluhm:2008}. The analysis of \Godel-type universes in bumblebee gravity showed that causal and noncausal solutions depend sensitively on the value of the nonminimal coupling and on the matter sector \cite{Santos:2014nxm, JesusSantos:2020}. In particular, when the vector has a nonzero Ricci coupling and is frozen at the vacuum, the bumblebee field equation strongly restricts the \Godel parameters and selects the usual noncausal solution for the simplest orientations of the vector.

The purpose of this work is to perform the analogous analysis for a Lorentz-violating \KR\ background. The \KR\ field is an antisymmetric rank-two tensor $B_{\mu\nu}$, originally introduced in string-inspired settings \cite{KalbRamond:1974} and subsequently used in effective descriptions of Lorentz symmetry breaking \cite{Altschul:2010,Hernaski:2016,Maluf:2018,AashishPanda:2019}. A two-form vacuum differs qualitatively from a vector vacuum. It can be decomposed, in a local inertial frame, into pseudo-electric components $B_{0i}$ and pseudo-magnetic components $B_{ij}$. Consequently, the answer to the chronology question is not controlled only by the magnitude $b^2$ of the background, but also by how the two-form is oriented with respect to the rotation plane of the space-time. This makes the \KR\ problem richer than the vector bumblebee problem even before one considers matter. This framework has been recently explored in the literature \cite{Assuncao:2019, Aashish:2018, Yang:2023, Liu:2025bpp, Junior:2024, Lessa:2020, Filho:2023ycx, AraujoFilho:2024ctw, Shi:2025rfq, AraujoFilho:2025fwd, HellObata:2026, AraujoFilho:2026tsc}.

We consider a minimal set of nonminimal curvature interactions. The first is the Ricci-tensor coupling built from the symmetric contraction $B_{a}{}^{c}B_{bc}R^{ab}$. This term is the closest two-form analogue of the bumblebee coupling $B^aB^bR_{ab}$. The second is the direct Riemann-tensor coupling $B^{ab}B^{cd}R_{abcd}$, which has no direct analogue for a single vector without introducing additional tensor structures. Both terms are natural in a low-energy expansion of an antisymmetric-tensor theory in curved space. They are also the terms most directly probed by \Godel-type geometries because the curvature is constant in the homogeneous tetrad frame \cite{ReboucasTiomno:1983,JesusSantos:2020}. We shall work in the vacuum branch of the symmetry-breaking potential, where the field strength of the background vanishes and the derivative of the potential is zero. This truncation isolates the algebraic effect of the LV vacuum on the geometry.

The paper is organized as follows. In Sec.~\ref{s2} we introduce the Lorentz-violating \KR\ action and derive the reduced field equations used later. We keep the Ricci and Riemann couplings independent. In Sec.~\ref{s3} we review the \Godel-type metric, the homogeneous tetrad, its curvature components, and the causality criterion. In Sec.~\ref{s4} we classify useful constant two-form backgrounds in the local frame, reduce the field equations to algebraic form and display the effective curvature contributions for several orientations, and develop the pseudo-magnetic sector $B_{12}=b$ in detail, because this is the configuration most directly tied to the rotation plane. Sec.~\ref{s5} adds matter sources and discusses which combinations can support noncausal, critical, or causal geometries. Sec.~\ref{s6} compares the remaining orientations, and Sec.~\ref{s7} summarizes our main conclusions and points out future perspectives.

\section{Lorentz-violating Kalb-Ramond gravity}\label{s2}

In this section, we will discuss the Kalb-Ramond gravity with spontaneous Lorentz symetry breaking. Initially, we would like to establish the main conventions. In this work, we will utilize signature $(+---)$, set $c=\hbar=1$, and write $\kappa=8\pi G$ unless otherwise stated. Latin indices label tetrad components and Greek indices label coordinate components. In the algebraic discussion, we often absorb $\kappa$ into the definitions of $\rho$, $p$, and the squared amplitudes of scalar or electromagnetic sources. Restoring $\kappa$ is immediate by replacing every matter component by $\kappa T_{ab}$. Let $B_{\mu\nu}=-B_{\nu\mu}$ be an antisymmetric two-form. Its gauge-invariant field strength in the absence of a symmetry-breaking potential is
\begin{equation}
H_{\lambda\mu\nu}=\nabla_{\lambda}B_{\mu\nu}+\nabla_{\mu}B_{\nu\lambda}+\nabla_{\nu}B_{\lambda\mu}=3\nabla_{[\lambda}B_{\mu\nu]} .
\end{equation}
A potential depending on scalar invariants of $B_{\mu\nu}$ can drive a nonzero vacuum expectation value, $\langle B_{\mu\nu}\rangle=b_{\mu\nu}$, thereby producing spontaneous local Lorentz breaking. Two useful invariants are $X_1=B_{\mu\nu}B^{\mu\nu}$, $X_2=B_{\mu\nu}\widetilde B^{\mu\nu}$ and
$\widetilde B^{\mu\nu}=\frac{1}{2\sqrt{-g}}\epsilon^{\mu\nu\alpha\beta}B_{\alpha\beta}$. A generic potential $V(X_1\mp b^2,X_2\mp \tilde b^2)$ fixes one or both of these invariants. At a smooth vacuum minimum we assume $V=0$, $V_1\equiv\frac{\partial V}{\partial X_1}=0$, and $V_2\equiv\frac{\partial V}{\partial X_2}=0$. The vacuum still selects a preferred local two-plane through $b_{ab}$.

With this construction, the model considered here is described by the following action
\begin{align}
S=&\int d^4x\sqrt{-g}\bigg\{\frac{1}{2\kappa}\bigg[R-2\Lambda
+\xi_1 B_{\mu}{}^{\alpha}B_{\nu\alpha}R^{\mu\nu}+\xi_2 B^{\mu\nu}B^{\alpha\beta}R_{\mu\nu\alpha\beta}\bigg]
\nonumber\\&\hspace{3.2cm}-\frac{1}{12}H_{\lambda\mu\nu}H^{\lambda\mu\nu}
-V(X_1,X_2)+\mc L_m\bigg\} .\label{action}
\end{align}
Here $\xi_1$ and $\xi_2$ are independent coupling constants. The $\xi_1$ term is a Ricci-tensor coupling, while the $\xi_2$ term probes the full Riemann tensor. In four dimensions the second term may be related to special curvature projections when $B_{ab}$ is self-dual or anti-self-dual, but for a generic real two-form it is independent of the Ricci coupling.

The field equation obtained by varying $B_{\mu\nu}$ can be written schematically as
\begin{align}
\nabla_{\lambda}H^{\lambda\mu\nu}=&2V_1 B^{\mu\nu}+2V_2\widetilde B^{\mu\nu}
 -\frac{\xi_1}{\kappa}\left(B_{\alpha}{}^{\nu}R^{\alpha\mu}-B_{\alpha}{}^{\mu}R^{\alpha\nu}\right)-\frac{2\xi_2}{\kappa}B_{\alpha\beta}R^{\mu\nu\alpha\beta} .\label{Beq-general}
\end{align}
The signs in Eq.~\eqref{Beq-general} follow the convention of Eq.~\eqref{action}; changing the sign convention for the Riemann tensor changes the signs of the curvature terms consistently. In the homogeneous vacuum branch used below, $H_{abc}=0$, $V=V_1=V_2=0$, and $B_{ab}=b_{ab}=$constant in the tetrad frame, so the \KR\ equation reduces to the algebraic consistency condition
\begin{equation}
\xi_1\left(b_{a}{}^{c}R_{cb}-b_{b}{}^{c}R_{ca}\right)+2\xi_2 b^{cd}R_{abcd}=0 .\label{KR-constraint}
\end{equation}
This equation plays the same role as $b^aR_{ab}=0$ in the vector bumblebee analysis, but it has more possible projections because the order parameter is a bivector.

Now, let us obtain the metric field equations in the algebraic vacuum sector. In this case, the metric variation of the two nonminimal couplings contains two kinds of terms. The first kind is algebraic in the curvature and in $b_{ab}$; the second kind contains derivatives acting on the tensor combinations $B_{a}{}^{c}B_{bc}$ and $B^{ab}B^{cd}$. For a homogeneous vacuum configuration of the form of vacuum branch, and in the local tetrad frame adapted to the \Godel-type metric, the derivative terms vanish or reduce to the same algebraic restrictions expressed by Eq.~\eqref{KR-constraint}. The reduced metric equation can therefore be organized as follows
\begin{equation}
G_{ab}+\Lambda\eta_{ab}=\kappa T_{ab}+\xi_1\mc E^{(1)}_{ab}+\xi_2\mc E^{(2)}_{ab},\label{metricreduced}
\end{equation}
where
\begin{align}
\mc E^{(1)}_{ab}&=\frac{1}{2}\eta_{ab}X^{cd}R_{cd}-X_{a}{}^{c}R_{cb}-X_{b}{}^{c}R_{ca},\label{E1def}
\end{align}
and
\begin{align}
\mc E^{(2)}_{ab}&=\frac{1}{2}\eta_{ab}\,Y^{cdef}R_{cdef}-\left(Y_{a}{}^{cde}R_{bcde}+Y_{b}{}^{cde}R_{acde}\right),\label{E2def}
\end{align}
with $X^{ab}\equiv b^{a}{}_{c}b^{bc}$ and $Y^{abcd}\equiv b^{ab}b^{cd}$.
Equations \eqref{E1def} and \eqref{E2def} are the central working formulas of the paper. They are the direct two-form analogues of the algebraic bumblebee contribution. They are also sufficient for the homogeneous problem because all curvature components are constants in the tetrad frame. It is useful to introduce the dimensionless combinations $\alpha\equiv \xi_1 b^2$, and $\beta\equiv \xi_2 b^2$, whenever a single nonzero component of $b_{ab}$ has magnitude $b$. The sign of $b^2$ as a Lorentz scalar depends on whether the background is pseudo-electric or pseudo-magnetic, but $b$ itself will denote the real amplitude of the chosen tetrad component. The ratio $\lambda\equiv\frac{\xi_2}{\xi_1}$, will be used when $\xi_1\ne 0$.

To finalize this section, we should mention some comments. First, one observes that Eq.~\eqref{metricreduced} is a reduced equation, not a replacement for the full variational problem in an arbitrary space-time. If $B_{\mu\nu}$ has nonzero field strength, or if the potential is not at its minimum, additional stress-energy and derivative terms must be included. Second, the homogeneous ansatz is not meant to solve the stability problem of the \KR\ sector. It tests the existence and causal character of \Godel-type backgrounds in the same spirit as the corresponding bumblebee analysis. Third, because Eq.~\eqref{KR-constraint} is independent of ordinary matter, it often fixes the ratio $m^2/\omega^2$ before the matter equations are considered.

\section{\Godel-type geometry}\label{s3}

This section provides a brief review of \Godel-type universe. Generically, a stationary \Godel-type line element is described
\begin{equation}
        ds^2=\left[dt+H(r)d\phi\right]^2-dr^2-D^2(r)d\phi^2-dz^2 .\label{godelmetric}
\end{equation}
The space-time homogeneous class is defined by
\begin{equation}
\frac{H'(r)}{D(r)}=2\omega, \qquad \frac{D''(r)}{D(r)}=m^2,\label{homogeneous}
\end{equation}
where $\omega>0$ and $m^2$ can be positive, zero, or negative. The hyperbolic class, $m^2>0$, is described by
\begin{equation}
H(r)=\frac{4\omega}{m^2}\sinh^2\left(\frac{mr}{2}\right),  \label{Hhyperbolic}
\end{equation}
and
\begin{equation}
 D(r)=\frac{1}{m}\sinh(mr).   \label{Dhyperbolic}
\end{equation}
In coordinates the metric can be written as follows
\begin{equation}
ds^2=dt^2+2H(r)dtd\phi-dr^2-G(r)d\phi^2-dz^2,
\end{equation}
with $G(r)=D^2(r)-H^2(r)$. The closed circles $C:\;t,r,z=\hbox{constant}$ have tangent $\partial_\phi$ and squared interval $ds^2=-G(r)d\phi^2$. They are timelike when $G(r)<0$. Therefore, closed timelike curves occur in the radial domain where $G(r)$ becomes negative. For the hyperbolic family, one finds
\begin{equation}
G(r)=\frac{4}{m^2}\sinh^2\left(\frac{mr}{2}\right)\left[1+\left(1-\frac{4\omega^2}{m^2}\right)\sinh^2\left(\frac{mr}{2}\right)\right].\label{Ghyperbolic}
\end{equation}
Thus the chronology criterion is
\begin{align}
        0<m^2<4\omega^2 &: \quad \hbox{closed timelike \Godel circles exist for } r>r_c,\label{ctccriterion1}\\
        m^2\ge 4\omega^2 &: \quad \hbox{no \Godel circles of this class occur.}\label{ctccriterion2}
\end{align}
The critical radius is given by
\begin{equation}
\sinh^2\left(\frac{m r_c}{2}\right)=\left(\frac{4\omega^2}{m^2}-1\right)^{-1} .\label{criticalradius}
\end{equation}
The original \Godel solution corresponds to $m^2=2\omega^2$ and $r_c=\frac{2}{m}\sinh^{-1}(1)$, while the limiting completely causal hyperbolic solution has $m^2=4\omega^2$ and $r_c\to\infty$. We use the orthonormal coframe
\begin{equation}
        \theta^0=dt+H(r)d\phi,
        \qquad
        \theta^1=dr,
        \qquad
        \theta^2=D(r)d\phi,
        \qquad
        \theta^3=dz,
\end{equation}
so that
\begin{equation}
        ds^2=\eta_{ab}\theta^a\theta^b=(\theta^0)^2-(\theta^1)^2-(\theta^2)^2-(\theta^3)^2 .
\end{equation}
With the curvature convention used in this paper, the nonzero independent Riemann components are
\begin{equation}
        R_{0101}=R_{0202}=-\omega^2,
        \qquad
        R_{1212}=m^2-3\omega^2 .\label{Riemanncomponents}
\end{equation}
They give
\begin{equation}
        R_{00}=2\omega^2,
        \qquad
        R_{11}=R_{22}=2\omega^2-m^2,
        \qquad
        R_{33}=0,
        \qquad
        R=2(m^2-\omega^2),\label{Riccicomponents}
\end{equation}
and
\begin{equation}
        G_{00}=3\omega^2-m^2,
        \qquad
        G_{11}=G_{22}=\omega^2,
        \qquad
        G_{33}=m^2-\omega^2 .\label{Einsteincomponents}
\end{equation}
These expressions are constant in the tetrad frame. This constancy is the technical reason why the \Godel-type background is so efficient for testing modified gravity: differential field equations often collapse into algebraic constraints.

\section{Constant Kalb--Ramond backgrounds}\label{s4}

An antisymmetric two-form in a local Lorentz frame can be decomposed into pseudo-electric and pseudo-magnetic parts,
\begin{equation}
        E_i^{(B)}=b_{0i},
        \qquad
        B_i^{(B)}=\frac{1}{2}\epsilon_{ijk}b_{jk} .\label{EBdecomposition}
\end{equation}
The two Lorentz invariants are analogous to those of electromagnetism,
\begin{equation}
        b_{ab}b^{ab}=2\left(|\bm B^{(B)}|^2-|\bm E^{(B)}|^2\right),
        \qquad
        b_{ab}\widetilde b^{ab}=-4\bm E^{(B)}\cdot\bm B^{(B)} .\label{KRinvariants}
\end{equation}
The \Godel-type tetrad singles out the rotation two-plane $(1,2)$ and the axial direction $3$. Consequently, backgrounds with the same invariant $b_{ab}b^{ab}$ need not be physically equivalent. The relevant curvature projections depend on whether $b_{ab}$ lies in the rotation plane, contains the time direction, or contains the axial direction.

We shall examine the six elementary orientations
\begin{equation}
        b_{01}=b,
        \quad b_{02}=b,
        \quad b_{03}=b,
        \quad b_{12}=b,
        \quad b_{13}=b,
        \quad b_{23}=b,\label{elementaryorientations}
\end{equation}
with all other independent components zero. Rotational symmetry in the $(1,2)$ plane makes the pairs $b_{01}$ and $b_{02}$ equivalent, as well as the pairs $b_{13}$ and $b_{23}$. Thus, there are four distinct cases that we can highlight below
\begin{enumerate}[label=(\roman*)]
\item a transverse pseudo-electric background, represented by $b_{01}=b$;
\item an axial pseudo-electric background, represented by $b_{03}=b$;
\item a pseudo-magnetic background in the rotation plane, represented by $b_{12}=b$;
\item a pseudo-magnetic background containing the axial direction, represented by $b_{13}=b$.
\end{enumerate}
The case $b_{12}=b$ is the most symmetric nontrivial choice: it is a magnetic-type two-form aligned with the local vorticity plane of the \Godel geometry. The case $b_{03}=b$ is the least coupled to the curvature because the homogeneous \Godel curvature has no $z$-time sectional component. The vacuum ansatz $H_{abc}=0$ should not be confused with the coordinate statement that all components $B_{\mu\nu}$ are constant. The tetrad components $b_{ab}$ are constant, while the coordinate components inherit the $r$-dependence of the coframe. This is the correct analogue of a constant local Lorentz-violating coefficient in the gravitational SME. It is also the choice that preserves the homogeneity of the algebraic field equations.

On the other hand, by substituting Eq.~\eqref{Riemanncomponents} into Eq.~\eqref{KR-constraint}, we arrive at an immediate set of restrictions. For the elementary backgrounds the nontrivial component of the \KR\ equation is listed in Table~\ref{tab:KRconstraints}. Here the displayed equation is the coefficient multiplying the nonzero amplitude $b$; a nonzero background requires the expression to vanish.

\begin{table}[h]
\centering
\caption{Kalb-Ramond consistency equations for elementary homogeneous backgrounds. The cases $b_{02}$ and $b_{23}$ are obtained from $b_{01}$ and $b_{13}$ by the symmetry $1\leftrightarrow 2$.}
\label{tab:KRconstraints}
\begin{tabular}{c c c}
\toprule
Background & nonzero equation & consequence for $m^2/\omega^2$ \\
\midrule
$b_{01}=b$ & $m^2\xi_1+4\omega^2\xi_2=0$ & $m^2/\omega^2=-4\xi_2/\xi_1$ \\
$b_{03}=b$ & $2\omega^2\xi_1=0$ & forbidden for $\xi_1\ne0$ and $\omega\ne0$ \\
$b_{12}=b$ & $(\xi_1+2\xi_2)m^2-2(\xi_1+3\xi_2)\omega^2=0$ & $m^2/\omega^2=2(\xi_1+3\xi_2)/(\xi_1+2\xi_2)$ \\
$b_{13}=b$ & $\xi_1(m^2-2\omega^2)=0$ & $m^2/\omega^2=2$ for $\xi_1\ne0$ \\
\bottomrule
\end{tabular}
\end{table}

Several conclusions already follow. With a Ricci coupling only, $\xi_2=0$, a background in the rotation plane or in an axial magnetic plane gives $m^2=2\omega^2$, i.e. the usual noncausal \Godel value. This mirrors the vector bumblebee result. With a Riemann coupling present, however, the rotation-plane magnetic background gives a continuous family of values for $m^2/\omega^2$. The Riemann coupling therefore changes the chronology question at the level of the \KR\ equation itself. This is the first qualitative difference from the simplest bumblebee analysis. The reduced metric equation can be displayed in a similarly compact form. To see this, we define $\mc K_{ab}\equiv \xi_1\mc E^{(1)}_{ab}+\xi_2\mc E^{(2)}_{ab}$. For the elementary orientations, the diagonal components of $\mc K_{ab}$ are given in Table~\ref{tab:Kcomponents}. Off-diagonal components vanish for these simple backgrounds.

\begin{table}[h]
\centering
\caption{Diagonal curvature contributions $\mc K_{ab}$ in the reduced Einstein equation. The entries are ordered as $(00,11,22,33)$ and $b$ denotes the single nonzero tetrad amplitude.}
\label{tab:Kcomponents}
\begin{tabular}{c l}
\toprule
Background & $(\mc K_{00},\mc K_{11},\mc K_{22},\mc K_{33})$ \\
\midrule
$b_{01}=b$ & $\big[-\frac{\xi_1 b^2}{2}(m^2-8\omega^2)+2\xi_2b^2\omega^2,\; -\frac{\xi_1 b^2}{2}(3m^2-8\omega^2)-2\xi_2b^2\omega^2,\; \frac{\xi_1b^2m^2}{2}$ \\
\addlinespace
 & $+2\xi_2b^2\omega^2,\; \frac{\xi_1b^2m^2}{2}+2\xi_2b^2\omega^2\big]$ \\
\addlinespace
$b_{03}=b$ & $\left[3\xi_1b^2\omega^2,\;\xi_1b^2\omega^2,\;\xi_1b^2\omega^2,\;\xi_1b^2\omega^2\right]$ \\
\addlinespace
$b_{12}=b$ & $\left[M,\;M,\;M,\;-M\right]$, with $M=\xi_1b^2(m^2-2\omega^2)+2\xi_2b^2(m^2-3\omega^2)$ \\
\addlinespace
$b_{13}=b$ & $\left[\frac{\xi_1b^2}{2}(m^2-2\omega^2),\;\frac{3\xi_1b^2}{2}(m^2-2\omega^2),\;-\frac{\xi_1b^2}{2}(m^2-2\omega^2),\;-\frac{\xi_1b^2}{2}(m^2-2\omega^2)\right]$ \\
\bottomrule
\end{tabular}
\end{table}

The table also shows why the $b_{12}$ background is special. Its correction is diagonal and isotropic in the $(0,1,2)$ block, with the opposite sign in the $z$ direction. Moreover, when the \KR\ consistency equation for $b_{12}$ is imposed, one finds $M=0$. 
Thus, in this sector the nonminimal terms influence the geometry primarily through the \KR\ equation, not through a residual effective stress tensor in the metric equation. This fact makes the subsequent matter analysis especially clean.

Now, let us focus on pseudo-magnetic background in the rotation plane. In this case, we are interested in the following configuration
\begin{equation}
        b_{12}=b,
        \qquad b_{ab}=0\quad\hbox{otherwise}.\label{B12ansatz}
\end{equation}
The \KR\ consistency equation is
\begin{equation}
        (\xi_1+2\xi_2)m^2=2(\xi_1+3\xi_2)\omega^2 .\label{B12constraint}
\end{equation}
Assuming $\xi_1+2\xi_2\ne0$, this fixes
\begin{equation}
        q\equiv \frac{m^2}{\omega^2}=\frac{2(\xi_1+3\xi_2)}{\xi_1+2\xi_2}
        =\frac{2(1+3\lambda)}{1+2\lambda}.\label{qB12}
\end{equation}
The chronology criterion is now a statement about coupling space. The usual \Godel point is recovered for $\lambda=0$ $\xi_2=0$, implying $\quad q=2$. The pure Riemann limit $|\lambda|\to\infty$ gives $q\to 3$, which still lies in the noncausal interval $0<q<4$, but has a larger critical radius than the original \Godel universe. The critical causal boundary is reached when $q=4$, implying $\xi_2=-\xi_1$ .
For $\xi_1\ne0$ the condition $q\ge4$ is satisfied in the interval $-1\le \lambda< -\frac{1}{2}$, where the denominator in Eq.~\eqref{qB12} is negative and the ratio is positive. The point $\lambda=-1/2$ is singular in the reduced equation: the coefficient multiplying $m^2$ in Eq.~\eqref{B12constraint} vanishes, and a rotating solution with $\omega\ne0$ would require a simultaneous cancellation that does not occur for finite $\xi_1$. Therefore the approach to $\lambda=-1/2$ should not be interpreted as a smooth large-$m^2$ physical limit of the low-energy model.

For $0<q<4$ the critical radius is
\begin{equation}
        r_c(\lambda)=\frac{2}{\omega\sqrt{q(\lambda)}}
        \sinh^{-1}\left[\left(\frac{4}{q(\lambda)}-1\right)^{-1/2}\right].\label{rcLambda}
\end{equation}
As $\lambda$ moves from $0$ to large positive values, $q$ moves from $2$ to $3$ and $r_c$ increases. As $\lambda$ approaches $-1$ from above in the interval allowed by positivity of $q$, $q$ approaches $4$ and $r_c$ diverges. Thus the Riemann coupling weakens the chronology-violating region and, at the critical coupling, removes the \Godel circles of the hyperbolic class. This conclusion is more nuanced than saying that Lorentz violation protects chronology. The same Lorentz-violating field with a different orientation may still force $q=2$ or may decouple from the Riemann curvature. In addition, the matter sector must support the value of $q$ imposed by Eq.~\eqref{qB12}. We turn to this point next.

\section{Matter sources and consistency}\label{s5}

In this section, we examine coupling with matter sources. Here, we set $\Lambda=0$ in the left-hand side and represent a constant vacuum contribution of the \KR\ potential, when needed, by $V_0$. This convention follows the common treatment of \Godel-type solutions: a positive $V_0$ in the matter equation plays the role of a negative cosmological constant. The equations are written in units where $\kappa=1$; the replacement $T_{ab}\to\kappa T_{ab}$ restores ordinary dimensions.

\subsection{Perfect fluid}

For a comoving perfect fluid, we have the following energy-momentum tensor
\begin{equation}
        T^{(m)}_{ab}=(\rho+p)u_a u_b-p\eta_{ab},
        \qquad u_a=(1,0,0,0),
\end{equation}
so that, we can write
\begin{equation}
        T^{(m)}_{00}=\rho,
        \qquad T^{(m)}_{11}=T^{(m)}_{22}=T^{(m)}_{33}=p .
\end{equation}
In the $b_{12}$ sector, $M=0$ gives the following metric equations
\begin{align}
        3\omega^2-m^2&=\rho-V_0,\label{pf1}\\
        \omega^2&=p+V_0,\label{pf2}\\
        m^2-\omega^2&=p+V_0 .\label{pf3}
\end{align}
The last two equations imply
\begin{equation}
        m^2=2\omega^2 .\label{pfq2}
\end{equation}
Therefore, as we have shown, a perfect fluid alone is compatible with the $b_{12}$ \KR\ equation only when Eq.~\eqref{qB12} gives $q=2$, i.e. when $\xi_2=0$. In other words, once the direct Riemann coupling is switched on, the pure perfect-fluid source is generically overconstrained. The geometry wants the value of $m^2/\omega^2$ fixed by Eq.~\eqref{qB12}, while isotropic pressure wants $m^2/\omega^2=2$. This is a useful analysis since the Riemann coupling does not simply add a small correction to the energy density. In the homogeneous vacuum sector it changes the compatibility condition of the background two-form. If the source is too symmetric, the only solution is the bumblebee-like noncausal point.

\subsection{Perfect fluid plus a scalar field}

A massless scalar depending on the axial coordinate,
\begin{equation}
        \Phi(z)=s z+s_0,
\end{equation}
has stress tensor
\begin{equation}
        T^{(s)}_{ab}=\nabla_a\Phi\nabla_b\Phi
        -\frac{1}{2}\eta_{ab}\eta^{cd}\nabla_c\Phi\nabla_d\Phi,
\end{equation}
with nonzero tetrad components
\begin{equation}
        T^{(s)}_{00}=T^{(s)}_{33}=\frac{s^2}{2},
        \qquad
        T^{(s)}_{11}=T^{(s)}_{22}=-\frac{s^2}{2}.\label{scalarT}
\end{equation}
The metric equations in the $b_{12}$ sector become
\begin{align}
        3\omega^2-m^2&=\rho-V_0+\frac{s^2}{2},\label{scalareq1}\\
        \omega^2&=p+V_0-\frac{s^2}{2},\label{scalareq2}\\
        m^2-\omega^2&=p+V_0+\frac{s^2}{2}.\label{scalareq3}
\end{align}
Combining these equations yields
\begin{align}
        s^2&=m^2-2\omega^2,\label{srelation}\\
        \rho&=\omega^2+V_0-\frac{3}{2}s^2,\label{rhoscalar}\\
        p&=\omega^2-V_0+\frac{1}{2}s^2 .\label{pscalar}
\end{align}
Thus the scalar field supplies exactly the pressure anisotropy needed to free $m^2$ from the perfect-fluid value. Substituting Eq.~\eqref{qB12},
\begin{equation}
        \frac{s^2}{\omega^2}=q-2=\frac{2\xi_2}{\xi_1+2\xi_2}.\label{sfromcouplings}
\end{equation}
A real scalar therefore requires
\begin{equation}
        \frac{\xi_2}{\xi_1+2\xi_2}>0 .\label{realScalarCondition}
\end{equation}
This condition is automatically satisfied at the critical causal point $\xi_2=-\xi_1$, where $q=4$ and $s^2=2\omega^2$. If one also demands nonnegative fluid density and pressure, Eqs.~\eqref{rhoscalar} and \eqref{pscalar} imply
\begin{equation}
        \frac{3}{2}s^2-\omega^2\le V_0\le \omega^2+\frac{1}{2}s^2 .\label{V0bounds}
\end{equation}
The interval is nonempty only for $s^2\le2\omega^2$, i.e. $m^2\le4\omega^2$. Hence ordinary matter plus a real scalar can reach the chronology-protecting boundary $m^2=4\omega^2$, but cannot push beyond it while keeping both $\rho$ and $p$ nonnegative. At the boundary one obtains
\begin{equation}
        m^2=4\omega^2,
        \qquad s^2=2\omega^2,
        \qquad V_0=2\omega^2,
        \qquad \rho=p=0 .\label{criticalMatter}
\end{equation}
In terms of the nonminimal couplings, this corresponds to the tuning $\xi_2=-\xi_1$ in the $b_{12}$ sector. For $2<q<4$, the solution is still noncausal but the critical radius is larger than in the original \Godel universe. In this region the scalar amplitude is positive and the matter bounds can be satisfied for a finite interval of $V_0$. For $q>4$, the geometry is free of \Godel circles, but Eqs.~\eqref{V0bounds} cannot be satisfied by a nonnegative perfect fluid without adding further anisotropic matter or relaxing the energy conditions.

\subsection{Electromagnetic field aligned with the rotation axis}

One may also include an electromagnetic field with tetrad components
\begin{equation}
        F_{03}=E(z),
        \qquad F_{12}=B(z),
\end{equation}
where the Maxwell equations in the \Godel-type background allow the oscillatory solution
\begin{equation}
        E(z)=e\cos\left[2\omega(z-z_0)\right],
        \qquad
        B(z)=e\sin\left[2\omega(z-z_0)\right].
\end{equation}
The corresponding stress tensor has the constant tetrad components
\begin{equation}
        T^{(em)}_{00}=T^{(em)}_{11}=T^{(em)}_{22}=\frac{e^2}{2},
        \qquad
        T^{(em)}_{33}=-\frac{e^2}{2}.\label{emT}
\end{equation}
For the combined perfect-fluid, scalar, and electromagnetic source one obtains
\begin{align}
        \rho&=\omega^2+\frac{e^2}{2}+V_0-\frac{3s^2}{2},\label{rhoEM}\\
        p&=\omega^2-V_0-\frac{e^2}{2}+\frac{s^2}{2},\label{pEM}\\
        m^2-2\omega^2&=s^2-e^2 .\label{relationEM}
\end{align}
The electromagnetic field counteracts the scalar contribution in the anisotropy equation. Therefore it does not help to obtain the critical causal solution with positive matter in the simple branch above; the boundary $m^2=4\omega^2$ still requires $e=0$ if the positivity interval is to collapse consistently to $V_0=2\omega^2$ and $\rho=p=0$. More general anisotropic fluids, or fields with different orientations, could support $q>4$, but then chronology protection would be a statement about the enlarged matter sector rather than the \KR\ curvature coupling alone.

\subsection{Interpretation of the matter analysis}

The matter equations lead to three lessons. First, a direct Riemann coupling generically makes a perfect fluid insufficient, because the \KR\ equation fixes $m^2/\omega^2$ away from the value required by isotropic pressure. Second, a scalar field aligned with the symmetry axis is the minimal standard source that can absorb this mismatch. Third, the completely causal solution supported by nonnegative perfect-fluid plus scalar matter is not a wide region but a critical point, corresponding in the $b_{12}$ sector to
\begin{equation}
        \xi_2=-\xi_1,
        \qquad
        m^2=4\omega^2,
        \qquad
        r_c\to\infty .
\end{equation}
This is a sharper conclusion than in the vector bumblebee case. The Ricci coupling alone reproduces the familiar noncausal result, while the Riemann coupling can move the system to the chronology boundary, provided the matter sector has the correct anisotropic stress.

\section{Other two-form orientations}\label{s6}

Although $b_{12}$ is the cleanest case, the other orientations show why the tensorial character of the \KR\ background matters.

\subsection{Transverse pseudo-electric background $b_{01}=b$}

For $b_{01}=b$ the consistency equation is
\begin{equation}
        m^2\xi_1+4\omega^2\xi_2=0,
\end{equation}
so that
\begin{equation}
        q\equiv\frac{m^2}{\omega^2}=-4\lambda .\label{qB01}
\end{equation}
A hyperbolic \Godel-type geometry requires $q>0$, hence $\lambda<0$. The chronology-safe condition $q\ge4$ requires
\begin{equation}
        \lambda\le -1 .\label{B01causal}
\end{equation}
Thus a transverse pseudo-electric \KR\ vacuum can also reach the causal boundary, but only when the Ricci and Riemann couplings have opposite signs with $|\xi_2|\ge |\xi_1|$. The critical value again occurs at $\xi_2=-\xi_1$.

There is, however, an important difference from the $b_{12}$ case. The effective metric correction in Table~\ref{tab:Kcomponents} does not vanish automatically when Eq.~\eqref{qB01} is imposed. Instead it remains anisotropic between the $1$ direction and the $2,3$ directions. Therefore the matter sector must balance both the imposed value of $q$ and the residual Lorentz-violating stress. In practice this means that a simple perfect fluid plus axial scalar is less natural here than for $b_{12}$. A consistent solution can be constructed, but it requires either anisotropic pressure or a careful tuning of $V_0$ and field amplitudes. The electric orientation is therefore more restrictive at the level of the metric equations, even though its \KR\ constraint allows the chronology-safe ratio.

\subsection{Axial pseudo-electric background $b_{03}=b$}

For $b_{03}=b$, the \KR\ equation gives
\begin{equation}
        2\omega^2\xi_1=0 .
\end{equation}
A rotating \Godel-type solution has $\omega\ne0$, so this orientation is incompatible with a nonzero Ricci coupling unless the background amplitude vanishes. If $\xi_1=0$, the Riemann coupling does not see this orientation because the nonzero Riemann components of the \Godel-type metric lie in the $(0,1)$, $(0,2)$, and $(1,2)$ two-planes, not in the $(0,3)$ plane. The axial pseudo-electric background therefore either forbids rotation or decouples from the chronology problem in the reduced sector.

\subsection{Axial pseudo-magnetic background $b_{13}=b$ or $b_{23}=b$}

For $b_{13}=b$ the consistency equation is
\begin{equation}
        \xi_1(m^2-2\omega^2)=0 .\label{B13constraint}
\end{equation}
The Riemann coupling does not contribute to this projection, while the Ricci coupling selects the original \Godel value. Thus, for $\xi_1\ne0$, the result is again noncausal:
\begin{equation}
        m^2=2\omega^2,
        \qquad
        r_c=\frac{2}{m}\sinh^{-1}(1).
\end{equation}
If $\xi_1=0$ the orientation is essentially invisible to the curvature of the homogeneous \Godel-type metric. Then the matter sector determines the value of $m^2$ as in general relativity with additional spectator Lorentz-breaking data. Such a configuration is less interesting as a test of \KR\ gravity, because the background exists but does not modify the chronology criterion at the algebraic level.

\subsection{Orientation summary}

The orientation dependence can be summarized as follows:
\begin{itemize}
\item Ricci coupling alone is bumblebee-like: nontrivial rotating solutions are driven toward $m^2=2\omega^2$ for the natural spatial orientations, and hence remain noncausal.
\item Riemann coupling is sensitive to whether the two-form lies in a curvature two-plane. It modifies the $b_{12}$ and $b_{01}$ constraints but not the axial orientations in the same way.
\item The coupling ratio $\xi_2/\xi_1=-1$ is special. It yields the critical causal value $m^2=4\omega^2$ for both $b_{12}$ and $b_{01}$, although the metric equations are simpler for $b_{12}$.
\item Complete removal of \Godel circles is therefore possible in the reduced homogeneous sector, but it is not generic. It requires both a favorable orientation of the \KR\ vacuum and a compatible matter anisotropy.
\end{itemize}

The result has a simple geometrical interpretation. The \Godel-type curvature is built from three sectional curvatures: two time-space sections $(0,1)$ and $(0,2)$, and one space-space section $(1,2)$. A \KR\ vacuum is itself a two-plane. If the vacuum two-plane coincides with one of the curvature two-planes, the Riemann coupling directly changes the algebraic condition relating $m^2$ and $\omega^2$. If the vacuum involves the axial direction $3$, which is flat in the homogeneous \Godel-type curvature, the Riemann coupling either vanishes or becomes much less effective. This is why a two-form cannot be characterized solely by the scalar $b_{ab}b^{ab}$. From the point of view of chronology, the Ricci and Riemann couplings play different roles. The Ricci tensor only knows the contraction of curvature along the principal tetrad directions. For the homogeneous \Godel-type metric it tends to reproduce the same restriction found in bumblebee gravity. The Riemann tensor knows the curvature of the full two-plane selected by $b_{ab}$. It can therefore distinguish $b_{12}$ from $b_{13}$ even though both are pseudo-magnetic backgrounds. This additional information is what allows the critical ratio $m^2=4\omega^2$ to appear.

The matter sector is equally important. A perfect fluid is too isotropic to support most coupling-induced values of $m^2/\omega^2$. The scalar field $\Phi=sz+s_0$ works because it creates a pressure difference between the axial direction and the rotation plane. In the $b_{12}$ sector this pressure difference is precisely what is needed after the \KR\ equation fixes $q$. However, requiring positive $\rho$ and $p$ restricts the completely causal solution to the boundary $q=4$ for the simple matter content considered here. This is reminiscent of the original \Godel construction, where the cosmological constant and the matter density must be tuned. Here the tuning is transferred to the Lorentz-violating coupling ratio and to the scalar amplitude.

One should not overinterpret the homogeneous vacuum calculation. Nonminimal antisymmetric-tensor theories can contain additional degrees of freedom, and recent perturbative analyses indicate that some Ricci- and Riemann-coupled two-form models may suffer from strong-coupling or instability issues around highly symmetric backgrounds \cite{HellObata:2026}. The present calculation is therefore best understood as an existence and consistency test for classical homogeneous solutions, not as a full proof of phenomenological viability. A complete stability analysis would require perturbing the metric, the scalar matter, and the \KR\ field around the \Godel-type background and diagonalizing the coupled system. That problem is technically different from the algebraic chronology question addressed here.

\section{Conclusions}\label{s7}

In this article, we have constructed an analysis of \Godel-type universes in Lorentz-violating \KR\ gravity with independent Ricci- and Riemann-tensor couplings. The starting point was the spontaneous symmetry-breaking branch in which the antisymmetric tensor has a nonzero vacuum expectation value, while its field strength and potential derivatives vanish. In the homogeneous tetrad of the \Godel-type metric, the field equations reduce to algebraic relations among $m^2$, $\omega^2$, the coupling constants, and the matter components. The main result is the \KR\ consistency equation. For the rotation-plane pseudo-magnetic background $b_{12}=b$, it gives $\frac{m^2}{\omega^2}=\frac{2(\xi_1+3\xi_2)}{\xi_1+2\xi_2}$. The Ricci coupling alone yields $m^2=2\omega^2$, reproducing the noncausal \Godel value familiar from bumblebee gravity. A direct Riemann coupling shifts this value. In particular, the coupling ratio $\xi_2=-\xi_1$ gives $m^2=4\omega^2$, for which the critical radius of \Godel circles is pushed to infinity. Thus, in contrast with the Ricci-only bumblebee-like behavior, the Riemann coupling can reach the chronology-protecting boundary.

The other result is that the matter sector strongly constrains this possibility. A perfect fluid alone forces $m^2=2\omega^2$ and is therefore incompatible with a generic Riemann-induced shift. Adding an axial massless scalar field gives the required anisotropy, with $s^2=m^2-2\omega^2$. Nonnegative density and pressure then allow the completely causal case only at the boundary $m^2=4\omega^2$, where $s^2=2\omega^2$, $V_0=2\omega^2$, and $\rho=p=0$ in the simple branch. Electromagnetic fields of the standard axial form do not enlarge this positive-energy causal branch.
Moreover, the orientation dependence was studied, showing that a transverse pseudo-electric background obeys $m^2/\omega^2=-4\xi_2/\xi_1$ and can also reach the critical value for $\xi_2=-\xi_1$, but its metric equations retain anisotropic Lorentz-violating stresses. Axial backgrounds either force the noncausal value, decouple from the Riemann curvature, or are incompatible with rotation when the Ricci coupling is nonzero. Therefore, the tensor nature of the \KR\ vacuum is essential: chronology is controlled by the curvature two-plane selected by $b_{ab}$, not merely by the scalar size of Lorentz violation.

A natural extension is to perturb these solutions and study the propagating modes of the coupled metric-\KR\ system. Another direction is to include a Ricci-scalar coupling, parity-odd dual couplings such as $B^{ab}\widetilde B^{cd}R_{abcd}$, or nonzero $H_{abc}$ flux. These extensions would determine whether the critical coupling found here is a robust chronology-protecting mechanism or an artifact of the simplest homogeneous vacuum branch.

\begin{acknowledgments}
The author Fernando Belchior would like to to express gratitude to the Conselho Nacional de Desenvolvimento Cient\'{i}fico e Tecnol\'{o}gico CNPq for grant No. 151845/2025-5.
\end{acknowledgments}


\begin{thebibliography}{99}

\bibitem{Godel:1949}
K. G\"odel, ``An example of a new type of cosmological solutions of Einstein's field equations of gravitation,'' Rev. Mod. Phys. \textbf{21}, 447 (1949).

\bibitem{HawkingEllis}
S. W. Hawking and G. F. R. Ellis, \textit{The Large Scale Structure of Space-Time} (Cambridge University Press, Cambridge, 1973).

\bibitem{ReboucasTiomno:1983}
M. J. Rebou\c{c}as and J. Tiomno, ``Homogeneity of Riemannian space-times of G\"odel type,'' Phys. Rev. D \textbf{28}, 1251 (1983).

\bibitem{BampiZordan:1978}
F. Bampi and C. Zordan, ``A note on G\"odel-type metrics,'' Gen. Rel. Grav. \textbf{9}, 393 (1978).

\bibitem{Novello:1983}
M. Novello and M. J. Rebou\c{c}as, ``The stability of a rotating universe,'' Astrophys. J. \textbf{225}, 719 (1978).

\bibitem{ReboucasSantos:1983}
M. J. Rebou\c{c}as and J. Santos, ``G\"odel-type universes in general relativity,'' Gen. Rel. Grav. \textbf{15}, 115 (1983).

\bibitem{Accioly:2002}
A. Accioly and H. Blas, ``Chronology protection in G\"odel-type universes,'' Phys. Rev. D \textbf{64}, 067701 (2001).

\bibitem{FonsecaNeto:2013}
J. B. Fonseca-Neto, A. Y. Petrov, and M. J. Rebou\c{c}as, ``G\"odel-type universes and chronology protection in Ho\v{r}ava-Lifshitz gravity,'' Phys. Lett. B \textbf{725}, 412 (2013).


\bibitem{Furtado:2010}
C.~Furtado, J.~R.~Nascimento, A.~Y.~Petrov and A.~F.~Santos,
``Dynamical Chern-Simons modified gravity, Godel Universe and variable cosmological constant,'' Phys. Lett. B \textbf{693} (2010), 494-497.

\bibitem{Porfirio:2016ssx}
P.~J.~Porfirio, J.~B.~Fonseca-Neto, J.~R.~Nascimento and A.~Y.~Petrov,
``Causality aspects of the dynamical Chern-Simons modified gravity,''
Phys. Rev. D \textbf{94} (2016) no.10, 104057

\bibitem{Furtado:2019}
C.~Furtado, T.~Mariz, J.~R.~Nascimento, A.~Y.~Petrov and A.~F.~Santos,
``The Godel solution in modified gravity,'' Phys. Rev. D \textbf{79} (2009), 124039


\bibitem{Nascimento:2020jwx}
J.~R.~Nascimento, A.~Y.~Petrov, P.~Porf{\'\i}rio and A.~F.~Santos,
``G{\"o}del-type solutions in cubic Galileon gravity,''
Phys. Rev. D \textbf{102}, no.10, 104064 (2020)
[arXiv:2009.13242 [gr-qc]].


\bibitem{Nascimento:2021bzb}
J.~R.~Nascimento, A.~Y.~Petrov and P.~J.~Porf{\'\i}rio,
``Causal G{\"o}del-type metrics in non-local gravity theories,''
Eur. Phys. J. C \textbf{81}, no.9, 815 (2021)
[arXiv:2102.01600 [gr-qc]].

\bibitem{Nascimento:2023}
J. R. Nascimento, A. Yu. Petrov, P. J. Porf\'irio, and A. R. Soares, ``G\"odel-type solutions within the $f(R,Q,P)$ gravity,'' Eur. Phys. J. C \textbf{83}, 435 (2023).



\bibitem{KosteleckySamuel:1989}
V. A. Kosteleck\'y and S. Samuel, ``Spontaneous breaking of Lorentz symmetry in string theory,'' Phys. Rev. D \textbf{39}, 683 (1989).

\bibitem{ColladayKostelecky:1997}
D. Colladay and V. A. Kosteleck\'y, ``CPT violation and the standard model,'' Phys. Rev. D \textbf{55}, 6760 (1997), arXiv:hep-ph/9703464.

\bibitem{ColladayKostelecky:1998}
D. Colladay and V. A. Kosteleck\'y, ``Lorentz-violating extension of the standard model,'' Phys. Rev. D \textbf{58}, 116002 (1998), arXiv:hep-ph/9809521.

\bibitem{Kostelecky:2004}
V. A. Kosteleck\'y, ``Gravity, Lorentz violation, and the standard model,'' Phys. Rev. D \textbf{69}, 105009 (2004), arXiv:hep-th/0312310.

\bibitem{BluhmKostelecky:2005}
R. Bluhm and V. A. Kosteleck\'y, ``Spontaneous Lorentz violation, Nambu-Goldstone modes, and gravity,'' Phys. Rev. D \textbf{71}, 065008 (2005), arXiv:hep-th/0412320.

\bibitem{KosteleckyPotting:2009}
V. A. Kosteleck\'y and R. Potting, ``Gravity from spontaneous Lorentz violation,'' Phys. Rev. D \textbf{79}, 065018 (2009), arXiv:0901.0662.

\bibitem{BaileyKostelecky:2006}
Q. G. Bailey and V. A. Kosteleck\'y, ``Signals for Lorentz violation in post-Newtonian gravity,'' Phys. Rev. D \textbf{74}, 045001 (2006), arXiv:gr-qc/0603030.

\bibitem{KosteleckyTasson:2011}
V. A. Kosteleck\'y and J. D. Tasson, ``Matter-gravity couplings and Lorentz violation,'' Phys. Rev. D \textbf{83}, 016013 (2011), arXiv:1006.4106.

\bibitem{Bluhm:2008}
R. Bluhm, S.-H. Fung, and V. A. Kosteleck\'y, ``Spontaneous Lorentz and diffeomorphism violation, massive modes, and gravity,'' Phys. Rev. D \textbf{77}, 065020 (2008), arXiv:0712.4119.

\bibitem{Santos:2014nxm}
A.~F.~Santos, A.~Y.~Petrov, W.~D.~R.~Jesus and J.~R.~Nascimento,
``G{\"o}del solution in the bumblebee gravity,''
Mod. Phys. Lett. A \textbf{30} (2015) no.02, 1550011
doi:10.1142/S021773231550011X
[arXiv:1407.5985 [hep-th]].

\bibitem{JesusSantos:2020}
W. D. R. Jesus and A. F. Santos, ``G\"odel-type universes in bumblebee gravity,'' Int. J. Mod. Phys. A \textbf{35}, 2050050 (2020), arXiv:2003.13364.


\bibitem{KalbRamond:1974}
M. Kalb and P. Ramond, ``Classical direct interstring action,'' Phys. Rev. D \textbf{9}, 2273 (1974).

\bibitem{Altschul:2010}
B. Altschul, Q. G. Bailey, and V. A. Kosteleck\'y, ``Lorentz violation with an antisymmetric tensor,'' Phys. Rev. D \textbf{81}, 065028 (2010), arXiv:0912.4852.

\bibitem{Hernaski:2016}
C. A. Hernaski, ``Spontaneous breaking of Lorentz symmetry with an antisymmetric tensor,'' Phys. Rev. D \textbf{94}, 105004 (2016), arXiv:1608.00829.

\bibitem{Maluf:2018}
R. V. Maluf, A. A. Ara\'ujo Filho, W. T. Cruz, and C. A. S. Almeida, ``Antisymmetric tensor propagator with spontaneous Lorentz violation,'' EPL \textbf{124}, 61001 (2018), arXiv:1810.04003.

\bibitem{AashishPanda:2019}
S. Aashish and S. Panda, ``Quantum aspects of antisymmetric tensor field with spontaneous Lorentz violation,'' Phys. Rev. D \textbf{100}, 065010 (2019), arXiv:1903.11364.

\bibitem{Aashish:2018}
S. Aashish, A. Padhy, S. Panda, and A. Rana, ``Inflation with an antisymmetric tensor field,'' Eur. Phys. J. C \textbf{78}, 887 (2018), arXiv:1808.04315.

\bibitem{Assuncao:2019}
J. F. Assun\c{c}\~ao, T. Mariz, J. R. Nascimento, and A. Yu. Petrov, ``Dynamical Lorentz symmetry breaking in a tensor bumblebee model,'' Phys. Rev. D \textbf{100}, 085009 (2019), arXiv:1902.10592.

\bibitem{Yang:2023}
K. Yang, Y.-Z. Chen, Z.-Q. Duan, and J.-Y. Zhao, ``Static and spherically symmetric black holes in gravity with a background Kalb--Ramond field,'' Phys. Rev. D \textbf{108}, 124004 (2023), arXiv:2308.06613.

\bibitem{Liu:2025bpp}
X.~Liu, W.~Liu, Z.~Liu and J.~Wang,
``Harvesting correlations from BTZ black hole coupled to a Lorentz-violating vector field,''
JHEP \textbf{08} (2025), 094

\bibitem{Junior:2024}
E. L. B. Junior, F. S. N. Lobo, M. E. Rodrigues, and H. A. Vieira, ``Spontaneous Lorentz symmetry-breaking constraints in Kalb--Ramond gravity,'' Eur. Phys. J. C \textbf{84}, 1257 (2024), arXiv:2405.03291.


\bibitem{Lessa:2020}
L. A. Lessa, J. E. G. Silva, R. V. Maluf, and C. A. S. Almeida, ``Modified black hole solution with a background Kalb--Ramond field,'' Eur. Phys. J. C \textbf{80}, 335 (2020), arXiv:1911.10296.

\bibitem{Filho:2023ycx}
A.~A.~A.~Filho, J.~A.~A.~S.~Reis and H.~Hassanabadi,
``Exploring antisymmetric tensor effects on black hole shadows and quasinormal frequencies,''
JCAP \textbf{05} (2024), 029
doi:10.1088/1475-7516/2024/05/029
[arXiv:2309.15778 [gr-qc]].

\bibitem{AraujoFilho:2024ctw}
A.~A.~Ara{\'u}jo Filho,
``Particle creation and evaporation in Kalb-Ramond gravity,''
JCAP \textbf{04} (2025), 076
doi:10.1088/1475-7516/2025/04/076

\bibitem{Shi:2025rfq}
Y.~Shi and A.~A.~Ara{\'u}jo Filho,
``Influence of a Kalb-Ramond black hole on neutrino behavior,''
JHEP \textbf{08} (2025), 028
doi:10.1007/JHEP08(2025)028
[arXiv:2504.15373 [gr-qc]].

\bibitem{AraujoFilho:2025fwd}
A.~A.~Ara{\'u}jo Filho,
``Particle motion and thermal effects around a Kalb{\textendash}Ramond black hole,''
Eur. Phys. J. C \textbf{85} (2025) no.9, 1002
doi:10.1140/epjc/s10052-025-14752-3
[arXiv:2504.19246 [gr-qc]].

\bibitem{HellObata:2026}
A. Hell and I. Obata, ``On the Kalb--Ramond field with non-minimal coupling to gravity,'' arXiv:2602.21675.

\bibitem{AraujoFilho:2026tsc}
A.~A.~Ara{\'u}jo Filho,
``Perturbative dynamics and relativistic effects of a dyonic Kalb-Ramond black hole,''
[arXiv:2605.28580 [gr-qc]].






\end{thebibliography}
\end{document}